\documentclass[prf,longbibliography,onecolumn,superscriptaddress,amsmath,amssymb,aps,floatfix]{revtex4-1}

\usepackage{graphicx}
\usepackage{dcolumn}
\usepackage{bm}
\usepackage[bbgreekl]{mathbbol}
\usepackage{overpic}
\usepackage{dcolumn}
\usepackage{epstopdf}
\usepackage{siunitx}
\usepackage{mathrsfs}
\makeatletter
\def\amsbb{\use@mathgroup \M@U \symAMSb}
\makeatother
\expandafter\let\csname equation*\endcsname\relax
\expandafter\let\csname endequation*\endcsname\relax
\usepackage{amsmath}

\newcommand{\ve}[1]{\boldsymbol{#1}}
\renewcommand{\vec}[1]{\boldsymbol{#1}}
\newcommand{\ma}[1]{\ensuremath{\mathbb{#1}}}

\newcommand{\tr}{{\rm Tr}}

\usepackage{xcolor}

\begin{document}

\title{Helical Ribbons: Simple Chiral Sedimentation}

\author{Elias Huseby}
\affiliation{Department of Physics, Wesleyan University, Middletown, CT 06459, USA}
\author{Josephine Gissinger}%
\affiliation{%
CNRS, IUSTI, Aix Marseille University, Marseille, France}%
\author{Fabien Candelier}%
\affiliation{%
CNRS, IUSTI, Aix Marseille University, Marseille, France}%
\author{Nimish Pujara}%
\affiliation{%
Department of Civil and Environmental Engineering, University of Wisconsin-Madison, Madison, WI 53706 USA}%
\author{Gautier Verhille}%
\affiliation{%
Aix Marseille Univ, CNRS, Centrale Med, IRPHE, Marseille, France}%
\author{Bernhard Mehlig}%
\affiliation{%
Department of Physics, Gothenburg University, 41296 Gothenburg, Sweden}%
\author{Greg Voth}
\email{gvoth@wesleyan.edu}
\affiliation{Department of Physics, Wesleyan University, Middletown, CT 06459, USA}

\begin{abstract}

We study the design of chiral particle shapes that couple translation to rotation in viscous fluid flow.
We identify helical ribbons as particles with strong coupling whose symmetry ensures that the
centers of mass, buoyancy, resistance, and mobility coincide.  Co-centered particles such as helical ribbons provide an essential step in the hierarchy of increasingly complex translation-rotation coupling.  
Experiments on sedimenting helical ribbons measure  both relevant mobility tensors and show excellent agreement with simulations of ribbons made of interacting spheres. We observe quasi-periodic angular dynamics causing complex spatial trajectories.
In tilt-spin phase space, orbits are closed due to time-reversal and reflection symmetry. As the helical ribbon length increases, a bifurcation occurs at which the stable sedimentation orientations switch.  The particle at the bifurcation has axi-symmetric translation-rotation coupling and particularly simple sedimentation dynamics. Some special trajectories are unconfined in space and may contribute disproportionately to particle transport rates.

\end{abstract}

\maketitle

\section{Introduction}
How can shapes be designed to optimize coupling of translation to rotation? 
Couplings of a vector displacement in space to a pseudo-vector representing rotation are strongly constrained by reflection symmetries, and so this is sometimes called chiral design~\cite{efrati_orientation-dependent_2014}.  This problem appears in a wide range of fields including chiroptical response of molecules~\cite{crawford2006} or metamaterials~\cite{mun2020,ma2017}, chiral phases in crystals~\cite{hough2009} or nematics~\cite{dussi2016}, and the dynamics of colloids~\cite{krapf_chiral_2009} or larger particles~\cite{collins_lord_2021,Miara2024} moving through fluids.  Work on chiral design in fluid mechanics has often focused on propulsion in a single direction because many micro-organisims traverse their environments by rotating helical flagella~\cite{purcell1997flagellum,kim_macroscopic_2003,guasto2012,rodenborn_propulsion_2013,vach2015fast}, 
but in problems like sedimentation~\cite{doi_sedimentation_2005,krapf_chiral_2009}, environmental transport~\cite{sutherland2023}, and optimal trajectory planning for micro-robots~\cite{zhang}, particles encounter flows along all body coordinate directions and so the full three-dimensional translation-rotation problem must be understood.

As experiments have become available that measure full 3D translation-rotation coupling in fluids, it has become clear that we do not have an adequate roadmap to guide design of translation-rotation coupling in 3 dimensions. Miara {\em et al}~\cite{Miara2024} find chiral trajectories in sedimentation of bent discs, showing that the intuitively attractive idea that reflection symmetry prevents coupling of translation to rotation is wrong. Collins {\em et al}~\cite{collins_lord_2021} show that the isotropic helicoid, which can be seen as the simplest chiral geometry, has much smaller translation-rotation coupling than expected~\cite{Kelvin,Gustavsson2016}. Why do new particle shapes continue to surprise us when the underlying Stokes flow theory~\cite{Happel,Kim,Witten} is well known? And why has it been so difficult to develop a general understanding of the relationship between geometry and translation-coupling that can be transferred between different chiral design problems?  A central reason is that we do not have a clear roadmap that leads from shapes with simple translation-rotation coupling through to the general case. For non-chiral shapes in fluids, the roadmap starts with spheres and proceeds through spheroids to triaxial ellipsoids, spanning all possible drag tensors. But for translation-rotation coupling, no such hierarchy is known. 

For a particle moving through a viscous fluid with no velocity gradients, the particle velocity $\ve v$ and rotation rate $\ve \omega$ depend linearly on externally applied force $\ve f $ and torque $\ve \tau$:
\begin{equation}
\begin{bmatrix} \ve v \\ \ve \omega \end{bmatrix}  = \frac{1}{\mu}
\begin{bmatrix}\ma a' & {\ma b'}^{\sf T}  \\\ma b' & \ma c'  \end{bmatrix}
\begin{bmatrix}\ve f \\\ve \tau\end{bmatrix}
\label{eqft}
\end{equation}

The mobility tensors $\ma a'$, $\ma b'$, and $\ma c'$ are determined by particle shape and orientation~\cite{Happel}, with primes indicating that these are in the laboratory frame.
In sedimentation problems, the torque can be computed about a unique point at which $\ve \tau$ is zero~\cite{Witten}. Then with the rotation matrix from the body-fixed frame to the lab frame, \ma R, the equations of motion are
\begin{subequations}
\label{eq:eom}
\begin{eqnarray}
\tfrac{{\rm d} }{{\rm d}t}{\ve x} &=& \ve v\,, \quad \mu \ve v = 
\ma R \; \ma a\; \ma R^{-1} \ve f \label{eq:vf} \\
\tfrac{{\rm d} }{{\rm d}t} \ma R&=& \ve \omega \wedge \ma R\,,\quad \mu \ve \omega = 
\ma R \; {{\ma b}}  \; \ma R^{-1} \ve f \,.\label{eq:wf}
\end{eqnarray}
\end{subequations}

The full 3D design problem has 18 distinct parameters in $\ma a$, $\ma b$, and $\ma c$ which vary non-linearly as a function of particle geometry, independent of the applied force or torque. \ma a  is symmetric and can be specified by 6 parameters. Three of these parameters can be ignored by working in the \ma a eigenframe. \ma b has 9 parameters.  \ma c is also symmetric and so has 6 parameters.   
For the sedimentation case where the only external interaction is gravitational, it is possible to set the applied torque to zero and the number of parameters can be reduced to 12.  
Particles generally have distinct centers of mass and buoyancy, so in general there is an applied gravitational torque. However, there is a uniquely defined center where the torque due to these two forces cancel.  In the neutrally buoyant case, this center will be at infinity.    Defining torques around this point allows the sedimentation problem to be handled without external torques, and so the \ma c tensor can be ignored, although it requires the \ma b tensor to be computed around a center that is determined by the mass distribution and not simply geometry.   

\begin{figure}
\centering
    \includegraphics[width=15cm]{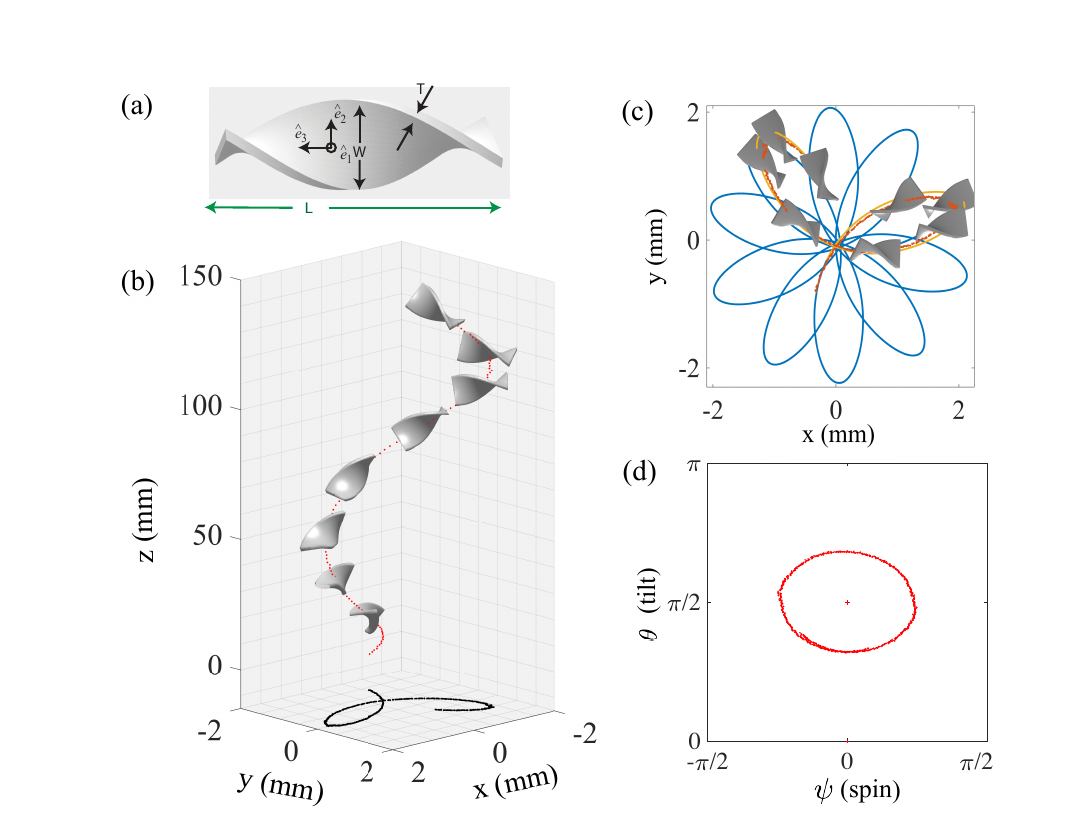}
  \caption{ \label{traj} (a) Body coordinate system and dimensions for a helical ribbon with Length $2\pi L/s=5\pi/4$ (b) A measured trajectory in space (Helix is not to scale, length $3\pi/4$)  (c) top view. Red dots are measurements. Yellow line is a numerical integration using measured parameters for this particle.  Blue line continues the numerical trajectory.  (d) Same measured trajectory in tilt-spin space.  Supplemental video is available~\cite{sm}.
}
 \end{figure}

To make progress toward a unified understanding of the relationship between geometry and dynamics, we need simpler classes of shapes. In this paper we identify the class of co-centered particles whose geometric symmetries ensure that the centers of mass, buoyancy, resistance, and mobility coincide, reducing the number of parameters from 18 to 9 (3 for each mobility tensor). Helical fibers, the chiral particles that have received the most attention in the literature~\cite{purcell1997flagellum,kim_macroscopic_2003,rodenborn_propulsion_2013,kramel_preferential_2016,palusa_sedimentation_2018, djutanta_decoding_2023}, are only co-centered for special lengths.  We identify helical ribbons (Fig.~\ref{traj}(a)) as chiral co-centered particles with strong translation-rotation coupling. 

Sedimentation is both a natural case for understanding the design of translation-rotation coupling and a problem with important applications including transport of plankton~\cite{guasto2012} and microplastics~\cite{sutherland2023}. Witten and Diamant~\cite{Witten} review the wide range of possible dynamics of chiral sedimentation. Particles typically have periodic orbits in orientation when \ma b is symmetric and   
approach fixed orientations when there is an anti-symmetric part of \ma b \cite{doi_sedimentation_2005,Moths2013}.  However, this picture is incomplete since periodic orbits are found in a case with anti-symmetric part of \ma b by Miara {\em et al}~\cite{Miara2024,Vaquero_2024}.
Analytic or numerical predictions for the \ma b tensor that are validated by experimental measurements are rare.~\cite{purcell1997flagellum,tozzi_settling_2011,collins_lord_2021,Miara2024}. The consequences of shape symmetries for particle dynamics have been explored in some detail~\cite{bretherton1962motion,krapf_chiral_2009,fries2017angular,collins_lord_2021,Witten,Ishimoto_2020a,Ishimoto_2020b}, but there still does not exist a general framework for classifying particle trajectories based on particle geometry, or for guiding design of particle geometry to obtain desired mobility tensors.

In this paper, we measure the 3D dynamics of sedimenting helical ribbons and extract full mobility tensors experimentally.  Numerical simulations show good agreement with the measured mobility tensors.  We find an unexpected bifurcation as a function of helical ribbon length at which the ribbons have axi-symmetric translation-rotation coupling and are particularly simple co-centered particles. The spatial trajectories of helical ribbons form beautiful spirograph patterns. Each co-centered particle is observed to have an infinite set of special initial orientations at which the particle position moves off to infinity rather than orbiting.

\section{Experimental Methods}

Helical ribbons are formed by a thin rectangular sheet twisted about its long axis so that its long edges form a double helix, shown in Fig.~\ref{traj}(a). Their three perpendicular $\pi$ rotational symmetries ensure that helical ribbons are always co-centered since any rotational symmetry fixes all centers to lie along the rotation axis. We fabricated particles with projection micro stereolithography 3D printing 
in order to obtain precise geometry and good density homogeneity.  Some earlier methods we tried for fabricating particles are described in Appendix~\ref{app:exp}. Four helical ribbons were used with lengths labeled by their twist angles, $2 \pi L/s = 3 \pi /4$, 5$\pi$/4, 4$\pi$/3, and 3$\pi$/2, where $s=20$ mm is the length over which the helix twists through a full rotation. All particles had width W=4 mm, thickness $T=0.5$ mm, and density 1.20~g/cm$^3$.  
 They were dropped in a 20x20x20 cm tank of silicon oil with mass density $\rho=$0.97~g/cm$^3$ and kinematic viscosity $\mu/\rho$= 520 cSt. The particle Reynolds number based on length, $Re_p = L |\vec{v}| \rho/\mu$, is 0.027$<Re_p<$0.092, firmly in the viscous limit.

\begin{figure*}[t]
     \centering
     \includegraphics[width=15cm]{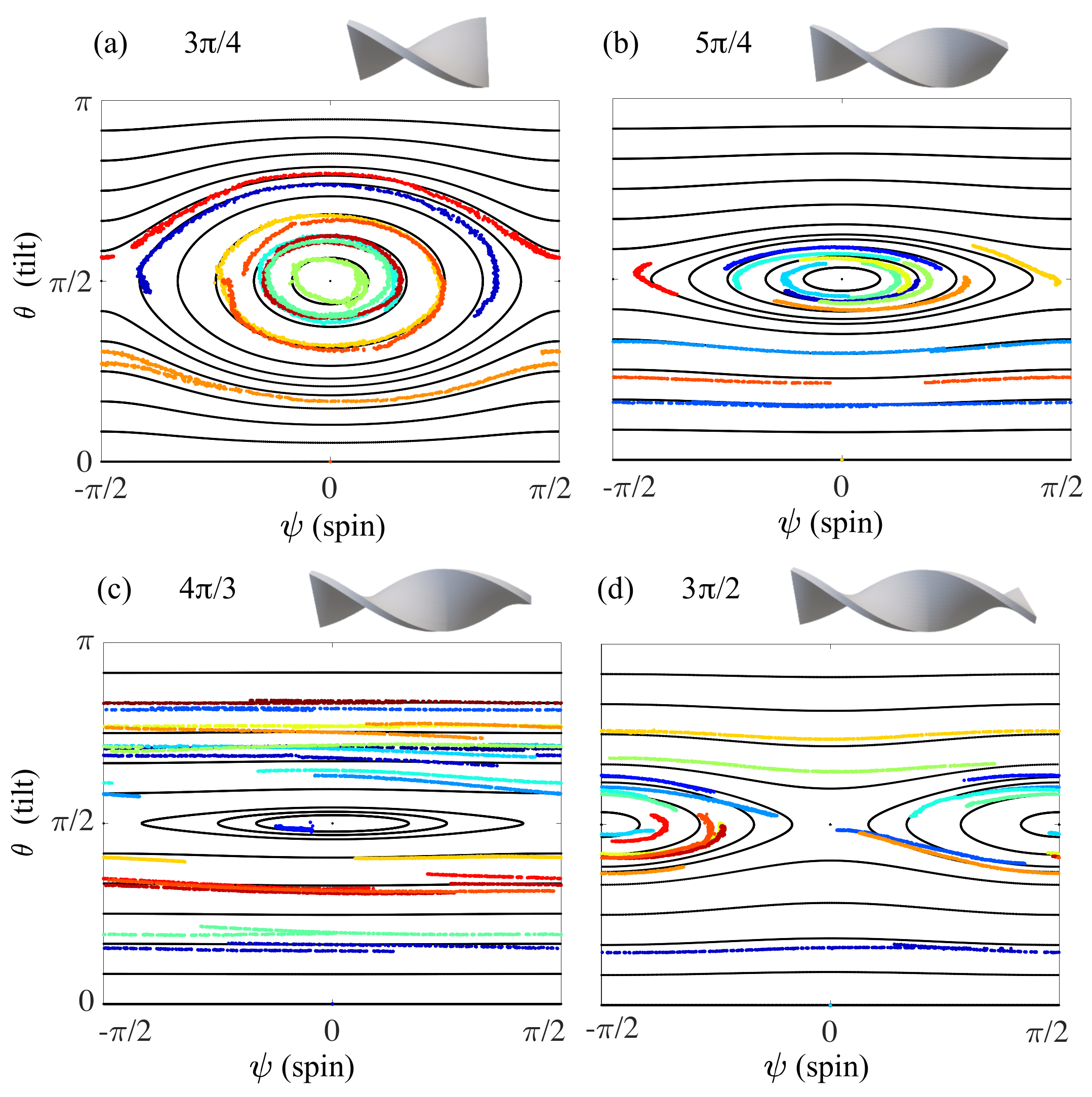}
  \caption{\label{phase} Angular dynamics for different particle lengths, $2\pi L/s$: (a) $3\pi/4$, (b) $5\pi/4$, (c) $4\pi/3$, and (d) $3\pi/2$. Particle images are above each plot.    Solid black lines show numerically integrated trajectories.  Color symbols indicate distinct experimental trajectories.}
 \end{figure*}

Three-dimensional positions and orientations of particles were reconstructed using 3 cameras with nearly orthogonal viewing angles~\cite{gautier,sm}. It is convenient to measure particle orientation using Euler angles~\cite{goldstein1980classical} where $\psi$ measures spin about the particle's long axis, $\theta$ is the polar angle measuring the tilt of body coordinate $\hat{\vec{e}}_3$ away from vertical, and $\phi$ is the azimuthal angle. We define $\psi = 0$ to be when $\hat{e}_1=\hat{e}_3\times\hat{z}$.
Note that tilts away from horizontal correspond with deviations from $\theta=\frac{\pi}{2}$.

Initial camera calibration parameters are obtained using a static two-plane calibration and then they are refined to sub-pixel accuracy with a dynamic calibration using images of small settling metal spheres.  To find the position and orientation of the helical ribbon from video images, we project a set of points on the surface of an ideal 3D helical ribbon onto the image planes of the three cameras using the camera calibration. The average distance of each these points from a bright pixel in the thresholded image is minimized to find the 3 position and 3 orientation coordinates.

The range of helical ribbon lengths we study and the distance over which we track them are limited by experimental constraints.  
Camera resolution prevents us from extending the tracking to large distances since we need about 50 pixels across a particle to accurately measure orientation.   For long particles, the period of closed orbits becomes very large and so we can only track them over a small portion of a closed orbit.  For particles much shorter than $2 \pi L/s=3 \pi/4$,  there is not enough structure in the bright field images to reliably measure orientations. Limitations of 3D printer resolution makes it difficult to scale down the size of the particles.

\section{Results}
Figure~\ref{traj}(b) shows a typical helical ribbon trajectory. Seen from above [Fig.~\ref{traj}(c)], the trajectory forms a quasi-periodic orbit with lobes similar to those made by a spirograph toy.   In the space of tilt ($\theta$) and spin ($\psi$) shown in Fig.~\ref{traj}(d), it traces a simple closed periodic orbit.  These closed orbits are similar to those observed for bent disks~\cite{Miara2024}.   The mechanism for tilt-spin orbits in helical ribbons is that a tilted ribbon drifts diagonally and begins to spin. The spinning changes which end of the ribbon experiences greater drag and so the tilt changes. A video of the example trajectory shown in Fig. 1 is available at: {\scriptsize https://www.youtube.com/watch?v=n9FWTLHiUK4}. The angular phase space is pictured at the bottom left with the same axes as in Fig. 1(d).

Figure~\ref{phase} shows many trajectories measured with different initial orientations for each of the four particles.  From these data, the six unknown eigenvalues of \ma a and \ma b can be determined using least-squares fitting of the experimental position and orientation data to trajectories numerically integrated using Eqs.~\ref{eq:eom}.  The black lines in Fig.~\ref{phase} show simulated trajectories using best fit mobility tensors. We see that theory and experiment agree  quite well. There are some differences, likely the result of imperfect particle geometry, foreign objects such as lint or bubbles attached to the particles, or systematic orientation measurement errors. 

For all four particles, and for all measured initial orientations, the trajectories appear to follow closed orbits in the $\theta$-$\psi$-plane, when periodic boundary conditions are considered.  Two fixed points, one a center and the other a saddle, appear in each phase diagram. A surprising feature of Fig.~\ref{phase} is that 
as the particle increases length from Fig.~2(c) to 2(d), there is a bifurcation at which the fixed points switch stability. At the bifurcation, the eigenvalues of the Jacobian both go through zero and all points along $\theta=\pi/2$ become fixed points.

To study the bifurcation as a function of ribbon length,  Eqs~\ref{eq:eom} can be used to obtain evolution equations for the Euler angles. For co-centered particles, we can use the reference frame in which $\ma a$ and $\ma b$ are both diagonal to obtain (in non-dimensional form)
\begin{subequations}
\label{eq:eqnsofmotion}
\begin{align}
\dot \psi&=  [(b_{11}-b_{22}) \cos^2\psi-b_{11}+b_{33}]\cos\theta\,,\\
\dot \theta &= (b_{11}-b_{22}) \sin\psi\cos\psi\sin\theta \,,\\
\dot\phi & = -(b_{11}-b_{22})\cos^2\psi +b_{11}\,.
\end{align}
\end{subequations}

Here the dynamics of the spin ($\dot{\psi}$) and tilt ($\dot{\theta}$) angles do not depend on the azimuthal angle $\phi$ because there is continuous rotational symmetry about the gravity direction. The continuous translational symmetry in space ensures that the angular dynamics do not depend on the particle's position, so the phase space is two dimensional.

Eqs.~\ref{eq:eqnsofmotion} show that the angular dynamics change as $b_{11}-b_{22}$ switches sign, because $\dot\theta$ vanishes. 
Linear stability analysis of Eqs.~\ref{eq:eqnsofmotion} confirms that for $b_{11}-b_{22}>0$, the fixed point at $\psi=0$, $\theta=\pi/2$ is a center while the fixed point at $\psi=\pi/2$, $\theta=\pi/2$ is a saddle, and they switch when $b_{11}-b_{22}$ changes sign.
Helicoids with $b_{11}=b_{22}$ have axisymmetric translation-rotation coupling and are apparently special particles with simpler dynamics than the general helical ribbon.

Eqs.~\ref{eq:eqnsofmotion} have a constant of motion~\cite{Witten}.
It can be obtained by integrating ${\rm d}\theta/{\rm d}\psi = \dot\theta/\dot\psi$:
\begin{equation}
\label{eq:result}
C= [{b_{33}-b_{11} + (b_{11}-b_{22})\cos^2\psi}]\sin^2\theta \,.
\end{equation}
This constant of motion, $C$,  exists because the dynamics are reversible \cite{arnold1984reversible,Strogatz,Politi1986,einarsson2016tumbling}, invariant under joint application
of time reversal and reflection.   In the phase space for helical ribbons, there are three lines of reflection, $\psi \rightarrow -\psi$,  $\psi- \pi/2 \rightarrow \pi/2-\psi$, and $\theta - \pi/2 \rightarrow \pi/2 - \theta$. Physically, these correspond to reflections across planes through the common centers of mass and mobility whose normal vectors are the particle's principle axes.  A physical reflection changes the sign of the eigenvalues of \ma b since it is a pseudo-tensor, which means that any convention defining geometric handedness will be inverted.  From Eqs.~\ref{eq:eqnsofmotion}, we see that the reflection causes all time derivatives to switch sign, confirming that co-centered particles always have reversible dynamics.   Particles with an anti-symmetric part of \ma b do not have reversible dynamics in general and tend to converge to orientations given by the eigenvectors of \ma b~\cite{Witten}.  Cases such as the bent disk, with an anti-symmetric part of \ma b and closed orbits, arise due to different physical symmetries~\cite{Miara2024} but still have reflection-time reversal symmetry in phase space.

Note that the angular dynamics in Eqs.~(\ref{eq:eqnsofmotion}) do not conserve phase-space volume which would globally ensure closed orbits in two dimensional phase space.  Reversibility implies closed orbits only for trajectories that twice crosses a line of reflection symmetry~\cite{Strogatz}.  For co-centered particles with three lines of reflection symmetry in phase space, every trajectory has two crossings and is closed.

\begin{figure}[b]
     \centering
     \includegraphics[width=15cm]{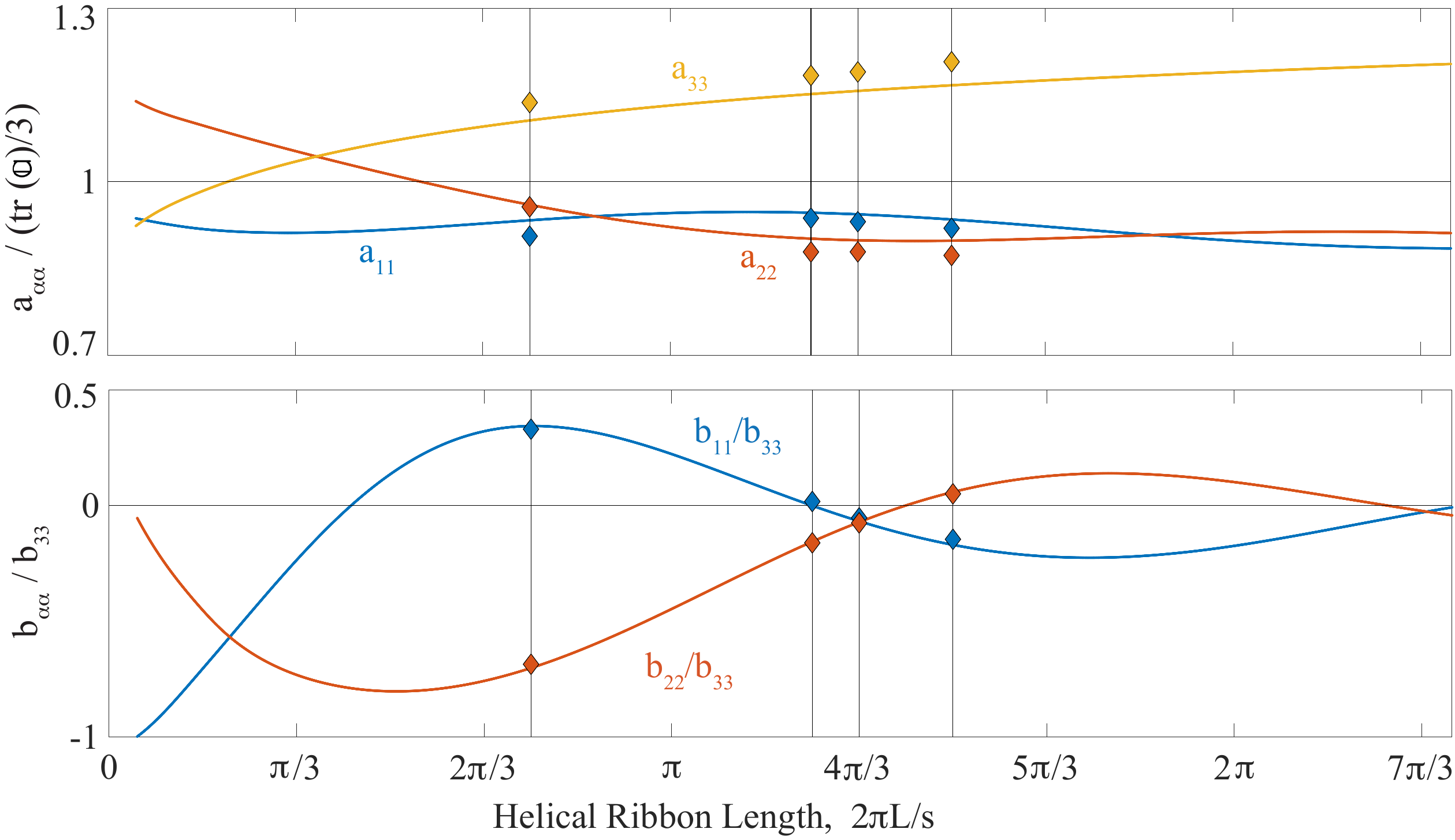} 
       \caption{ Diagonal elements of the mobility tensors for \ma a (top) and \ma b (bottom) as functions of helical ribbon length. Diamonds indicate experimentally measured values. Solid lines are numerical simulations using the bead-model described in the text.}
       \label{tensorcomp}
 \end{figure}

  \begin{figure}
     \centering
     \includegraphics[width=11cm]{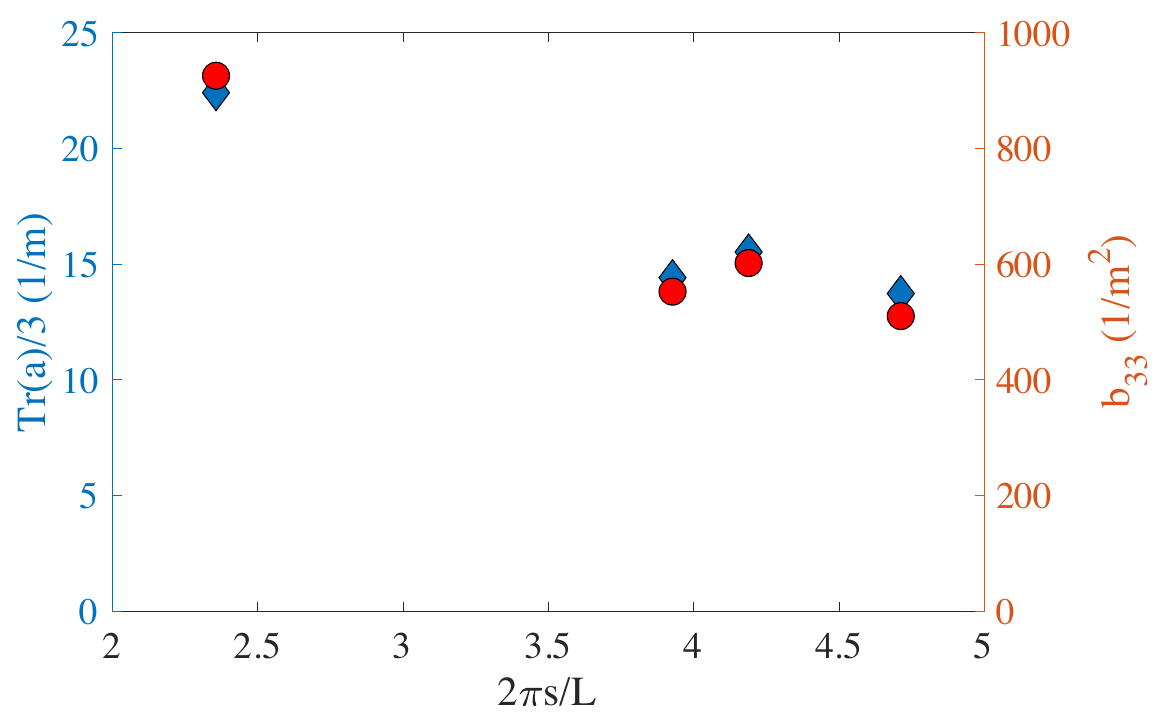}
 \caption{Normalization factors for mobility tensors as a function of helical ribbon length.  (Blue) Average of the eigenvalues of \ma a   (Red) Largest eigenvalue of \ma b.}
     \label{supp_norm}
 \end{figure}

 \begin{figure*}[t]
     \centering
     \includegraphics[width=14cm]{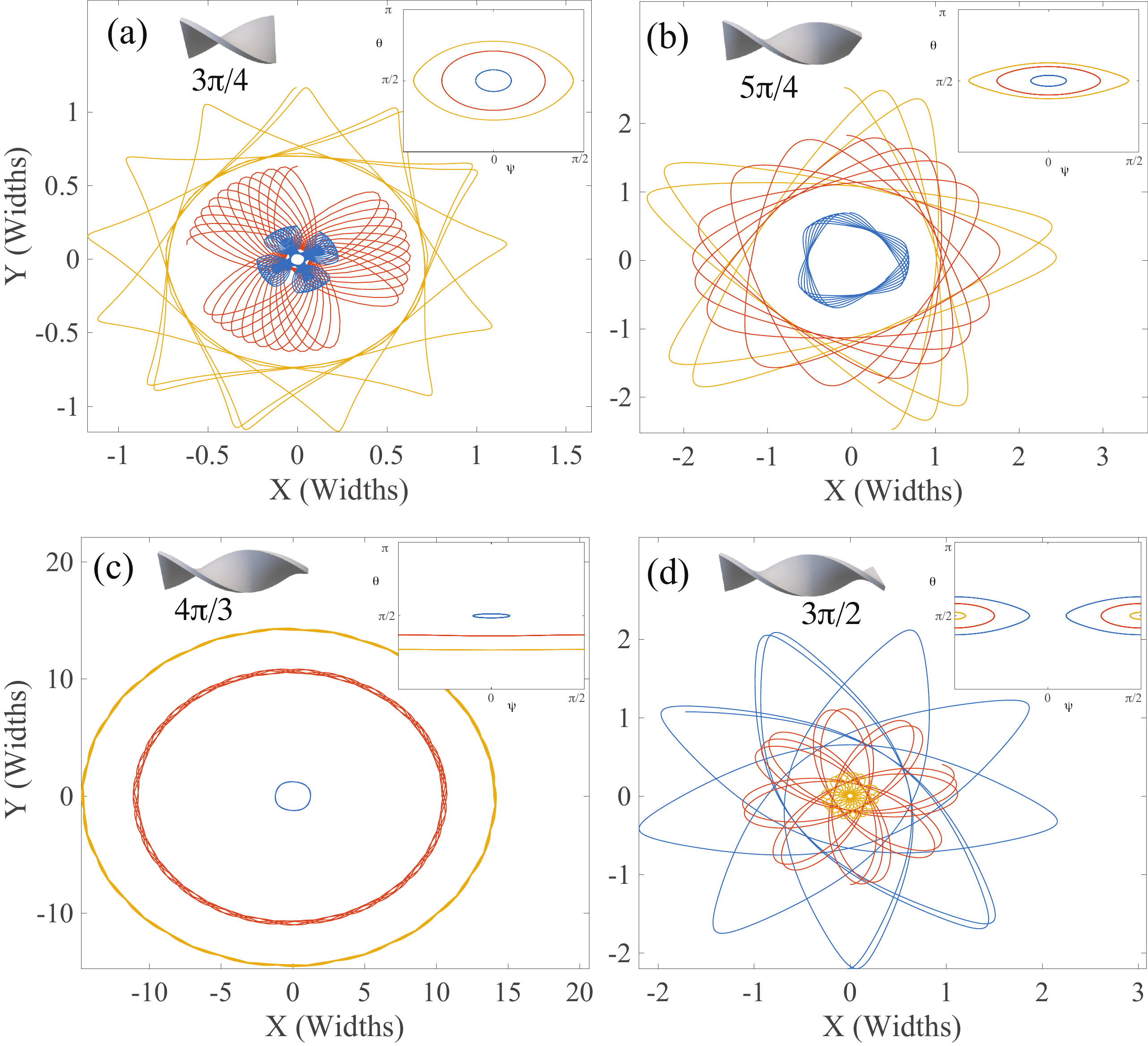}
  \caption{\label{spatial}  Spatial trajectories from numerical simulations for selected initial conditions.  Insets show the same trajectories (matched by color) in tilt-spin phase space.   Helical ribbons lengths are (a) $3\pi/4$, (b) $5\pi/4$, (c) $4\pi/3$, and (d) $3\pi/2$. Videos showing the full range of trajectories are available~\cite{sm}.}
\label{trajectories}
 \end{figure*}


To understand the possible mobility tensors for helical ribbons, we computed the mobility tensors numerically for ribbons made out of small hydrodynamically interacting spheres that are rigidly connected to each other~\cite{durlofsky1988dynamic,collins_lord_2021}. 
Normalized \ma a eigenvalues change little with length (Fig. 3(a)), except at very short lengths where the ribbon becomes more like a flat rectangle. The key is in the eigenvalues of \ma b shown in Fig. 3(b). Here the critical bifurcation lengths at which $b_{11}=b_{22}$ are clearly shown. In addition to the bifurcation just above $4\pi/3$ seen in Fig.~\ref{phase}, the numerical results show that there are two further bifurcations, near lengths 0.7 and 7.2.  There is no apparent special geometric symmetry for particles with any of these lengths. 
We expect the oscillation to continue as ribbons become longer and approach the simple chiral rod with only $b_{33}$ non-zero.

  \begin{figure}[b]
     \centering
     \includegraphics[width=13cm]{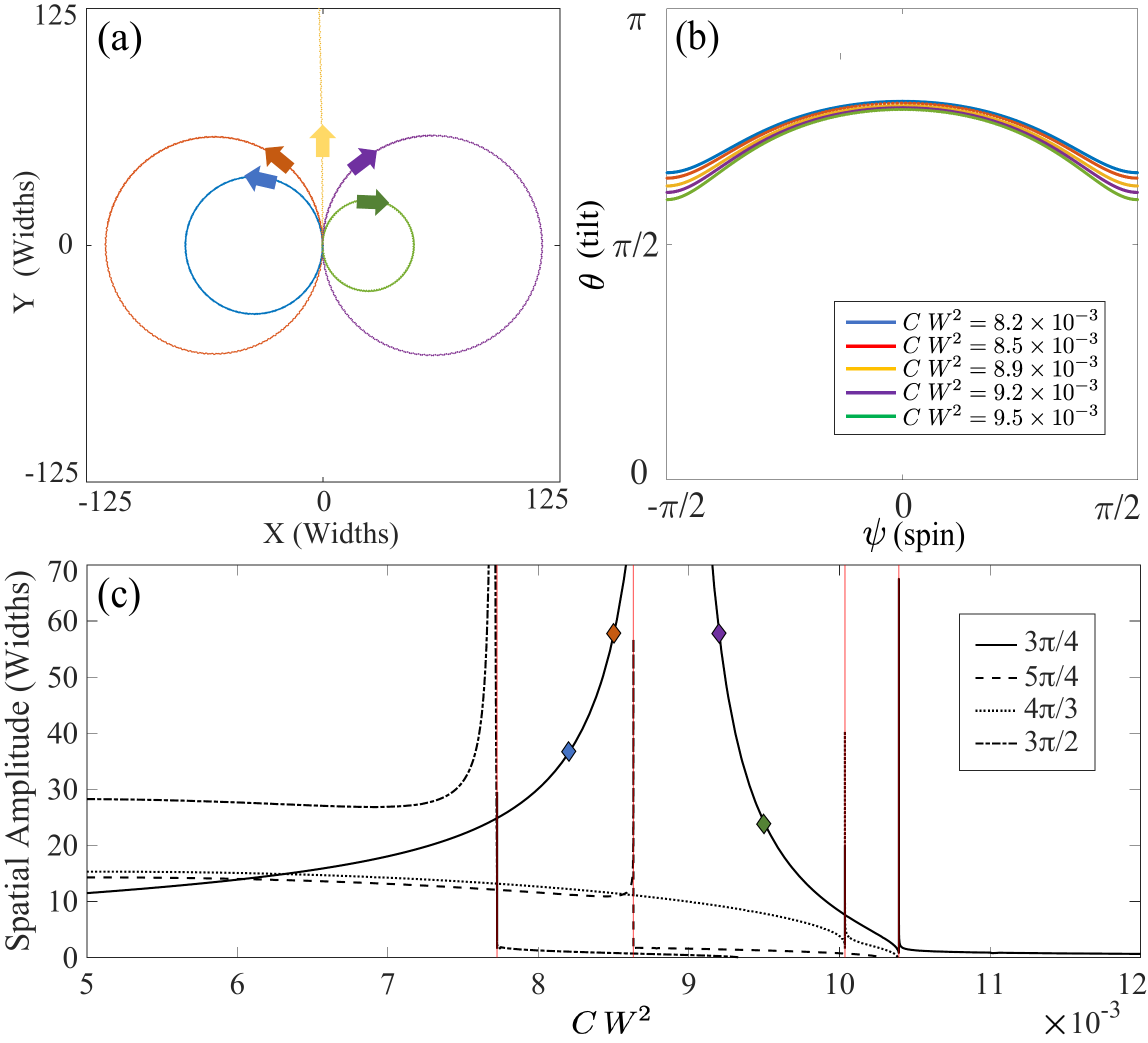}
  \caption{\label{curve} (a) Spatial trajectory curvature switching for a $3\pi/4$ particle. Arrows indicate direction on the simulated super-helical paths. (b) Phase space orbits of the trajectories in (a), matched by color. (c)  Maximum radius of the horizontal spatial motion for all each ribbon length as a function of the constant of motion $C$. Divergences representing spatially unconfined trajectories are seen in all cases. Red lines indicate the location of the separatrix that connects the two saddle points. Colored symbols show the trajectories in Figs. 5a and 5b.}
  
  \label{curvature}
 \end{figure}

Figure 3 also shows the experimentally measured eigenvalues of the $\ma a$ and $\ma b$ tensors. As normalized, the measurements are in quite good agreement with the computed values.  

In Fig. 3, the deviations between experiment and theory are slightly larger for \ma a. 
This may be a tank-size effect as confinement should especially decrease the measured $a_{11}$ and $a_{22}$, and the normalization could produce the observed discrepancy~\cite{roy2019symmetry}. We note that there is some dependence of the simulated mobility tensors on the number and size of the spheres used to represent the helical ribbon. This dependence mainly appears in the values of $\tr\, \ma a$ and $b_{33}$ and leaves the normalized eigenvalues unaffected.

We choose to normalize the elements of \ma a by tr(\ma a)/3 and \ma b by $b_{33}$. Physically, tr(\ma a)/3 is the average drag over all orientations. However, tr(\ma b) can be zero, so this is not a good normalization for \ma b. Instead, for this particle we choose to normalize by $b_{33}$, because it is larger than the other two eigenvalues and its sign is preserved as we change the helical ribbon length. In the limit of infinite length, $b_{33}$ is the only non-zero eigenvalue. 

Figure \ref{supp_norm} provides absolute scaling values of the normalization factors used for \ma a and \ma b.  We believe that small density differences between particles from different print batches causes the observed non-monotonic scaling. This effect appears as an uncertainty in the normalization factors but  does not significantly impact our conclusions.

The spatial trajectories produced by non-trivial \ma b tensors can be very intricate~\cite{Miara2024}. Orbits in $\theta$-$\psi$ phase space combine with precession in $\phi$ and the unequal eigenvalues of \ma a to produce motion at multiple frequencies. Fig.~\ref{spatial} shows spatial trajectories 
obtained numerically using fitted mobility tensor values.
Experimental trajectories are much shorter, but show the same structure. Trajectories that do not pass through $\theta=\frac{\pi}{2}$ (horizontal) remain tilted [Fig. \ref{trajectories}(c)] and typically have large super-helical spatial trajectories. Those that pass through $\theta=\frac{\pi}{2}$ have complex trajectories similar to those made by the spirograph toy [Figs. \ref{trajectories}(a,b,d)].  All four particles show both types, but we only show tilted trajectories in Fig. \ref{trajectories}(c).

To communicate the wide range of spatial trajectories that are possible, we created movies showing how the trajectories change as the initial orientation is continuously changed.  Movies are available at: {\scriptsize https://www.youtube.com/channel/UCVRg1jeDOSidH7oYnLdqZ2Q}. The movies show simulated trajectories using mobility tensors measured with fits to experimental data for each of the four helical ribbon lengths that we measured.

For co-centered helical ribbons, an interesting phenomenon occurs where the handedness of the spatial trajectory switches sign as the initial orientation is varied, shown in Fig.~\ref{curvature}.  In these cases there exists a critical trajectory for which the helical ribbon is not constrained within a finite horizontal area, in contrast to previous work~\cite{Makino}. These divergent trajectories may play a significant role in determining average transport statistics.

These divergences can occur in two cases: (1) when $b_{11}$ and $b_{22}$ have opposite signs, their contributions can cancel so that $\dot{\phi}$ integrated over a trajectory approaches zero, (2) for trajectories near the seperatrix, there is a tightly clustered infinite set of trajectories for which $\phi$ changes by integer multiples of $\pi$ during a period. 

The first mechanism is general to any shape with multi-signed \ma b eigenvalues and appears to have an impact over a substantial range of initial conditions (Fig. 6 (c)). The second mechanism, for trajectories near the seperatrix, guarantees an infinite number of spatially divergent trajectories for essentially all co-centered helical ribbons, but the divergences are narrow. In the videos linked above, we can observe (1) for the shortest and longest particles because the values of $b_{11}$ and $b_{22}$ (shown in Fig. 3) have opposite signs. However, the span of initial conditions over which (2) occurs is too small relative to the sampling resolution to be seen. 

\begin{figure}[tb]
     \centering
     \includegraphics[width=15cm]{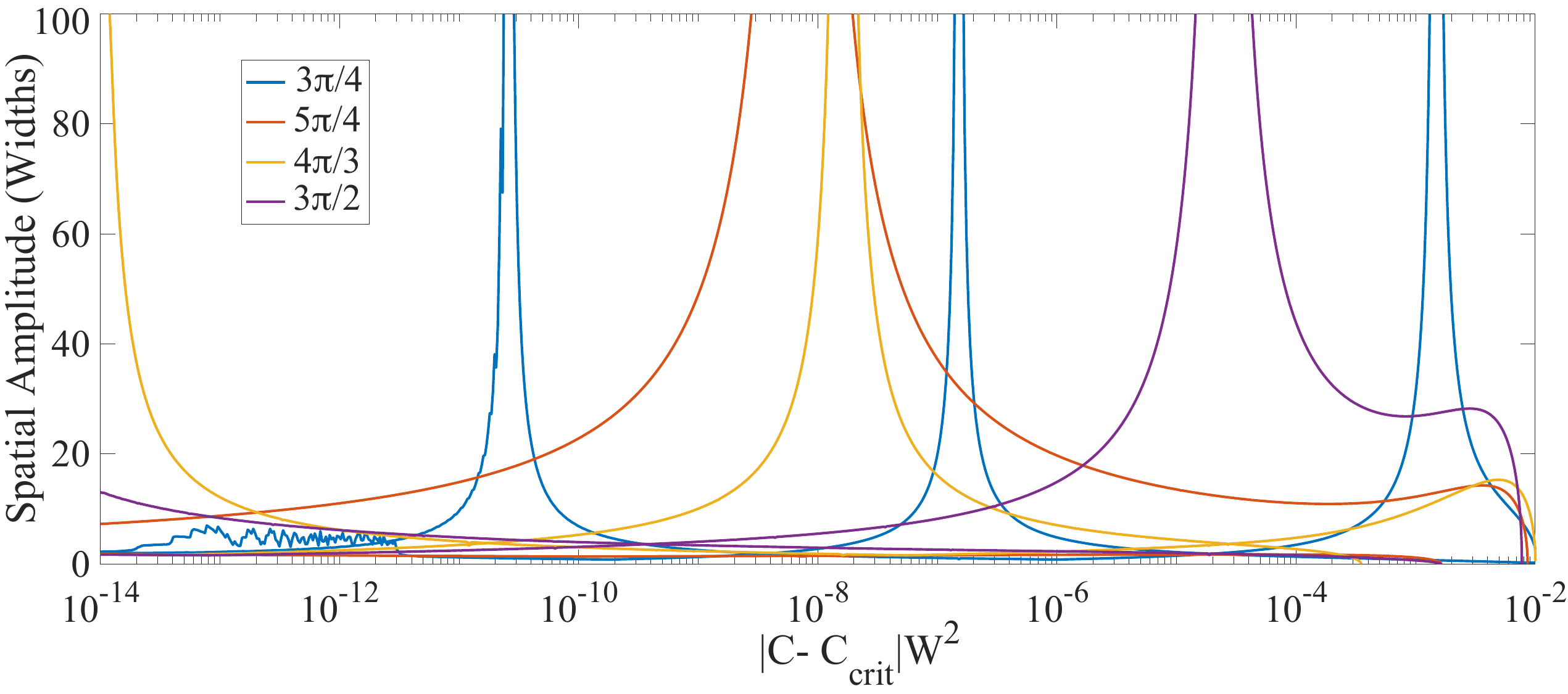}
 \caption{The spatial amplitude of simulated trajectories, defined as the radius of the circle that circumscribes the trajectory in the x-y plane, as a function of the difference of the conserved quantity $C$ from its critical value on the separatrix that connects the saddle points. Particle length of the four helical ribbons are labelled by color in the legend.}
     \label{supp_amp}
\end{figure}

Figure~\ref{supp_amp} shows the spatial amplitude of trajectories as a function of initial conditions quantified with $C$, similar to Fig. 6(c).  However, here we plot the difference between $C$ and $C_{\mathrm{crit}}$, the value of $C$ at the separatrix that connects the saddle points.  Plotted this way, the infinite sequence of different initial orientations for which the trajectories diverge can be seen. There are three places where $3 \pi/4$ particle shows divergences, two places for the $4 \pi/3$, and one for each of the others.  Since all but one of these represent azimuthal rotation by integer multiples of $\pi$ while the particles are near the saddle points, we expect there is an infinite sequence of such trajectories for all helical ribbons with a possible exception for special particles where $\dot{\phi}$ goes to zero at the saddle points. 

The switching of trajectory handedness for a single helical ribbon highlights the fact that no single number can quantify a particle's chirality;  instead, at least a second rank tensor is needed~\cite{efrati_orientation-dependent_2014}. The \ma b tensors measured here provide a useful general tool for quantifying the chiral coupling of a shape.

\section{Conclusion}

In conclusion, the helical ribbon is a chiral shape with 3 rotational symmetries that fix the centers of mass, buoyancy, mobility, and resistance to a single point. We find that helical ribbons exhibit strong translation-rotation coupling and identify them as a needed reference case with which more complex particles can be compared. With precise 3-dimensional measurements of position and orientation of helical ribbons sedimenting at particle Reynolds number less than 0.1, we measure both mobility tensors \ma a and \ma b. We observe angular orbits that are closed in the two-dimensional space of tilt and spin. 

There are helical ribbons with special lengths that have axisymmetric translation-rotation  coupling.  At these lengths, a bifurcation occurs where the fixed points in tilt-spin space switch between centers and saddles. These axisymmetric helicoids have a particularly simple \ma b tensor and can function as another step on the roadmap that extends to general helical ribbons with triaxial \ma b and on to yet more general shapes that are not co-centered.

Experiments that precisely fabricate complex shapes and measure full mobility tensors have now revealed several unexpected features of translation-rotation coupling during sedimentation~\cite{collins_lord_2021,Miara2024}.  Together with the present work, these point to an important design challenge to create geometries that optimize translation-rotation coupling.  Some progress in this direction has been made~\cite{keaveny_optimization_2013}, but the challenge goes far beyond optimizing propulsion efficiency in one dimension. For co-centered particles, each eigenvalue of \ma b can be optimized relative to the others to achieve different goals. A full picture of which geometries optimize translation-rotation coupling in Stokes flow could be a major step toward the broader goal of developing better ways to quantify and design chiral geometries.

\begin{acknowledgments}

We thank Bennet Grossman for assistance in setting up the experimental apparatus, Gleb Shevchuk for conversations about advanced 3D printing techniques and G. Sarnitsky for discussions.  We acknowledge support from the NSF under grant DMR-1508575 and CBET-2211704, the Army Research Office under grant W911NF-17-1-0176, the Initiative d’Excellence d’Aix-Marseille Université - A*MIDEX - AMX-19-IET-010, the French National Research Agency (ANR) in the framework of NetFlex Project (ANR-21-CE30-0040)  and Vetenskapsrådet under grant no. 2021-4452.

\end{acknowledgments}

\appendix

\section{Fabricating Co-Centered Particles}
\label{app:exp}

We tried three different methods for fabricating helical ribbons.  The first was mechanical twisting of polystyrene strips.  Strips of polystyrene sheet were cut on a paper cutter, twisted with a mechanical jig, annealed at 90$^o$ C, and then cut to the desired length.  This produced cost effective helical ribbons with variability in width, length, and pitch that required optical sorting to select specific particles.   There were also problems with non-reproducible sedimentation dynamics, likely a result of the thin polystyrene having mobility tensors that are very sensitive to tiny imperfections, lint, or bubbles.   

The second method was 3D printing with a Form2 printer from Formlabs.  This produced reproducible trajectories, but there was a measurable offset of the steady sedimentation orientation which we believe is due to density inhomogeneity in the printed material.  This method also has some challenges because it prints a single material and so supports must be removed by hand introducing some shape irregularities.    

We then obtained particles 3D printed with projection micro stereolithography developed by Boston Micro Fabrication whose equilibrium sedimentation orientation matches the expectations from the symmetry of the particles.  Their primary drawback is the higher cost per particle.  

We have come to view Stokes sedimentation dynamics as one of the most sensitive tools available for detecting irregularities in geometry or mass density of rigid bodies.


%

\end{document}